\documentstyle[12pt,fleqn]{article}
\def\Journal#1#2#3#4{{#1} {\bf #2}, #3 (#4)}
\def\NPB{{\em Nucl. Phys.} B}
\def\PLB{{\em Phys. Lett.} B}

\def\PR{\em Phys. Rev.}
\def\PRD{{\em Phys. Rev.} D}

\def\vep{\varepsilon}

\def\al{\alpha}
\def\be{\begin{equation}}
\def\ee{\end{equation}}
\def\bea{\begin{eqnarray}}
\def\eea{\end{eqnarray}}

\renewcommand{\thefootnote}{\fnsymbol{footnote}}

\begin{document}
\begin{titlepage}

\begin{center}
\large\bf Radiative Effects in Gauge Models with Homogeneous \\
Condensate and Curved Space-Time\footnotemark[1]
\end{center}
\vspace{0.5cm}
\noindent\footnotetext[1]{\noindent Talk presented at the Fourth International Workshop on
Quantum Theory under the Influence of External Conditions,
Leipzig University, Leipzig, September 1998.}
\begin{center}
V.~Ch.~Zhukovsky, V.~V.~Khudyakov, I.~V.~Mamsurov \\
{\sl Physical Faculty, Moscow State University,\\
119899, Moscow, Russia}
\end{center}

\begin{abstract}
Different models with nonabelian homogeneous condensate fields
are considered in the one-loop approximation. Effective action in a model
of gluodynamics in curved space is calculated. Free energy and its minimum
in a (2+1)-dimensional model of QCD are investigated.
Photon polarization operator (PO) is obtained.

\end{abstract}

\vspace{0.3cm}
\vfill
\end{titlepage}

\renewcommand{\thefootnote}{\arabic{footnote}}
\setcounter{footnote}{1}

\setcounter{page}{1}

\section{ Introduction}

Despite considerable efforts that have been undertaken in the recent
years there is no consistent theory of the QCD vacuum state at present.
Quark~\cite{Gell-Mann} and gluon condensates were for the first time
investigated in the framework of the sum-rule method by M.Shifman
et al.~\cite{shifman} and then basing upon some simple models of the gluon
background field (instantons~\cite{Shuryak,Reinhardt}, Savvidy
vacuum~\cite{savvidy}). Nonperturbative approaches were developed also
by Dosch and Simonov~\cite{Simonov} (cluster expansion of the vacuum
field method).
Active investigation of various radiative effects in non-Abelian external
fields has been initiated in this connection~\cite{[3_5]}.

On the other hand many problems of elementary particle physics have
thermodynamical aspects. Relevant to this are problems of phase transitions
in gauge theories at finite temperature and non-zero chemical
potential~\cite{[6]}, the question of possible stabilization of the classical
Yang-Mills field configurations on account of temperature
effects~\cite{[7],Eb_Vsh_Zhuk}, the problem of the particle energy radiative
shift and modifications of the particle electromagnetic properties on account
of finite temperature and density effects~\cite{[8_9]}, and a number of other
problems.
The chromomagnetic field configuration formed by nonabelian potentials
turned out to be stable with respect to the decay into pairs of real
particles~\cite{Brown,Agaev}. In this case substantially
nonperturbative dependence of such objects, as the photon PO, on the
vacuum field was obtained.

Recently much attention has been paid to the effects of topology~\cite{deser}
in forming the vacuum condensate, as well as to the non-analytic dependence
of the observables on the intensity of the gluon condensate~\cite{Ebert_Zh}.
The study of radiative effects in the 3-dimensional theories, carried out
recently by the MSU group~\cite{Kostya,Kolya} demonstrated importance of the
so called Chern-Simons topological term for regularization of infrared
divergences in calculating higher loop diagrams even in the case when
background fields are present.

\section{ One-loop effective action for gauge fields in curved space-time }

We investigate here a model of chromomagnetic vacuum in the space-time
continuum of nonzero curvature. Let us consider SU(2) gluodynamics in
{\bf S}$^2\times${\bf R}$^2$ space-time, which is a direct product of a plane
and a sphere of radius $\rho.$ Generating functional can be written in
the form
\be
  Z[\overline{A}, j]= \int\! da^a_\mu \:d\chi\:d \overline{\chi}
  \exp\left\{ -\int\!d^4x \sqrt{g}({\cal L} +ja) \right\},
\label{nq1} \ee
where potential $A^a_\mu= \overline{A}^a_\mu+ a^a_\mu,\;$
$\overline{A}^a_\mu$ is the background, $a^a_\mu$ are quantum fluctuations,
$
   {\cal L}= (F^a_{\mu\nu})^2/4 + (\overline{D}^{ab}_\mu a^\mu_b)^2/(2\xi)
   + \overline{\chi}_a (\overline{D}^2)_{ab}\chi_b\:.\:
$
Here $F^a_{\mu\nu}= \nabla_\mu A^a_\nu- \nabla_\nu A^a_\mu - ig(T^a)^{bc}
A^b_\mu A^c_\nu \;$ is the field tensor, $\;$ $\overline{D}^{ab}_\mu =
\delta^{ab}\nabla_\mu - ig(T^c)^{ab} \overline{A}^c_\mu $ is the background
derivative and $\chi_a$ are ghost fields.
Using $Z=\exp(\Gamma),\;$ one obtains effective action $\Gamma$
in the one-loop approximation
$
  \Gamma^{(1)}[\overline{A}]=
  \frac12 \mathop{\rm Tr}\ln \overline{\Theta}^{ab}_{\mu\nu} -
  \mathop{\rm Tr}\ln (-\overline{D}^2 )^{ab}
$ with the first and second terms as gluon and ghost contributions
respectively.

A model condensate abelian field can be chosen in the form
$\overline{A}^a_\mu = n^a\overline{A}_\mu,\;$
$\overline{F}^a_{\mu\nu}=n^a \overline{F}_{\mu\nu},\;$
$n^a=\mathop{\rm const},\;$ $n^a n^a = 1. $
Then, for $\xi = 1,$ upon diagonalizing $\overline{\Theta}^{ab}_{\mu\nu}$
one obtains $\overline{\Theta}^a_{\mu\nu} = - g_{\mu\nu}
(\nabla_\lambda - ig\nu^a \overline{A}_\lambda)^2 + ig\nu^a
\overline{F}_{\mu\nu} - R_{\mu\nu}. $ Here $\nu^a= \{1,1,0\}\; (a=1,2,3)$
are eigenvalues of matrix $n^cT^c.$

Consider large values of parameter $\chi^a= \omega^a\rho^2 \gg 1 \: $
($\omega^a=gH|\nu^a|).$
Define $p=[\sqrt{\chi}]$ as the integer part,
$\vep= \sqrt{\chi}-\rho$ as the fractional part.
Then in homogeneous chromomagnetic background field $H$ we obtain:
\be
  \mathop{\rm Im} \Gamma^{(1)}[\overline{A}]= -\frac{\Omega}{8\pi}(gH)^2
  (1-\frac2p\al);
\qquad
 \al=\left\{
   \begin{array}{ccc}
      -\vep &,& \vep<1/2 \\
      1-\vep &,& \vep>1/2
   \end{array}
 \right. ,
\ee
where $\Omega$ is the 4-volume. Since $2\al/\rho<1,$ we have
$\mathop{\rm Im} \Gamma^{(1)} \ne 0.$ After regularization the real part
of the effective action with account for the tachyonic mode contribution
reads:
\be
  \Gamma^{R\:(1)}[\overline{A}]=\frac{\Omega}{4\pi^2}(gH)^2\left(
  \frac{11}{12}\left(\ln\frac{gH}{\mu^2}-\frac12 \right)-
  \frac{\gamma-1}{p} \right),
\ee
where $\gamma$ is the Euler's constant.

\noindent
For small values of $\chi$ one obtains
$
  \mathop{\rm Im}\Gamma^{(1)}[\overline{A}]=-\Omega/(4\pi\rho^4),
$ and
\be
  \mathop{\rm Re}\Gamma^{(1)}[\overline{A}]=\frac{\Omega}{8\pi^2}(gH)^2
  \left( \frac{1}{\rho^2 gH}\ln\frac{\mu^2}{gH}
  -\frac{23}{12}\ln\frac{\mu^2}{gH}- \ln\rho^2 gH \right).
\ee
It should be emphasized that the imaginary part of the effective action,
$\mathop{\rm Im} \Gamma^{(1)},$ never disappears in this model and,
in contrast to~\cite{eli}, no stabilization occurs.

\section{ Free energy in (2+1)-dimensional SU(2) model of
          QCD with vacuum condensate }

Consider an SU(2) model of QCD in (2+1)-dimensional space-time at finite
temperature. The one-loop euclidean effective action
\be
  \Gamma^{(1)}=-\frac{1}{2}\int\frac{dq_4}{2\pi}\sum_r\ln(q_4^2+
  \vep_r^2(G))+ \sum_{j=1}^{N_f}\int\!\frac{dp_4}{2\pi}
  \sum_k\ln(p_4^2+\vep_k^2(Q_j))
\label{ea}\ee
is expressed in terms of one-particle boson $\vep_r(G)$ and fermion
$\vep_k(Q_j)$ spectra with quarks of color $a$ and flavor
$j=\overline{1,N_f}.$
Introducing finite temperature $T=1/\beta$ and generating functional $Z$
at $T\ne 0$ in the conventional way, we define the effective potential
as $V= -T\ln Z/L^2 =\Gamma^{(1)} L^{-3}.$ Here $L^3=\beta L^2 $
stands for the of 3-dimensional space-time volume. Separating the
background field energy density $V^{(0)}=(\overline{F }^a_{\mu\nu})^2/4$
and the one-loop quantum correction $v=v^G+v^Q,$ where $v^G$ and $v^Q$
are the quark and gluon contributions respectively, we have $V=V^{(0)}+v.$
Uniform vacuum condensate fields can be defined by gauge potential
$\overline{A}^a_\mu=\delta_{\mu 2}\delta_{a3}Hx_1 +
\delta_{\mu4}\delta_{a3}A_0 = \delta_{a3}\overline{A}_\mu\:.$ Chromomagnetic
field $H=\mathop{\rm const}$ and $A_0$-condensate are directed along third
color axis. The gluon energy spectrum with only physical degrees of
freedom taken into account (for zero chemical potential) is
$
  \vep(G)= \sqrt{2gH(n-1/2)}+gA_0 -i\epsilon, \; n=0,2,3,4,
  \ldots \,;
$  $\epsilon>0.$
Here $n=0$ corresponds to the tachyonic mode.

Substituting the energy spectrum in (\ref{ea}) with account for the
degeneracy of the energy spectrum we obtain the gluon contribution to the
effective potential, which is periodical in $gA_0$ with period $2\pi/\beta.$
Only those gauge transformations are admissible, for which ${\bf Z}_2$
symmetry is conserved
$A_0\to A'_0 =A_0+ 2\pi n/(\beta g)\;$ $(n\in {\bf Z}).$ It should be
noted that $v^G$ is real for $\sqrt{gH}< gA_0 <2\pi T-\sqrt{gH}.$
At $gA_0=\sqrt{gH}$ the effective potential has a singularity due to the
tachyonic mode contribution. The divergence is cured by account for radiative
correction
$\Delta\vep\sim \al_s=g^2/4\pi$ in the tachyonic energy
$\vep^2_{tach}=-gH-2i\al_s\sqrt{gH}.$
Temperature contribution can be separated as $v^G= v^G_{T=0}+ v^G_T. $
Here $v^G_{T=0}$ coincides with Trottier's result~\cite{trot}.
The effective potential reaches its extremum (minimum) value at
$ \sqrt{gH_0} \approx 0.218 g^2. $ For $H\to 0$ the temperature part reads
\be
  \left.v^G_T\right|_{H=0}
     \stackrel{T\gg gA_0}{\longrightarrow}
     -\frac{1}{\pi\beta^3} \left( \zeta (3)+\frac{(\beta gA_0)^2}{2}
  \left[ \ln(\beta gA_0)-\frac{3}{2} \right]\right).
\ee
In~\cite{trot} condensate $H$ was found to evaporate for $T>T_{cr}.$
However, the tachyonic mode was neglected there.
For convenience we introduce dimensionless variables
$x=\beta \sqrt{gH},\, y=\beta gA_0. $
Then the form of $V$ is demonstrated by the function
$U(x,y)= x^4 (T/g^2)^4+ u(x,y).$
For $T<T_{cr}\sim g^2$ the temperature contribution is small and $U(x,y)$
attains a global minimum at $x=x_{min}$ and $y=y_{min}.$
The second kind phase transitions in $U(x,y)$ occur with $y_{min}=\pi$
and $y_{min}=0$ interchanging. For $T>T_{cr}$ the values of $y_{min}$ and
$x_{min}$ decrease from $\pi$ to $0$ with growing temperature.

Polyakov loop is defined as
${\cal P}={\cal T}\exp\left[ i\int_0^\beta\!dt A_0^a \lambda^a/2 \right].$
If in the fundamental representation $\mathop{\rm Tr}_F({\cal P})=0, $ then
there exists a confinement phase. In our case
$\mathop{\rm Tr}_F({\cal P})= 2\cos(\beta gA_0/2).$
Therefore $\mathop{\rm Tr}_F({\cal P})=0$ is satisfied at $\beta gA_0=\pi.$

The quark contribution to the effective potential possesses a trivial
global minimum only. However, the quark contribution is much smaller than
the gluon one. Thus, quarks do not change the qualitative result given
by gluons.

We emphasize that the tachyonic modes are responsible for instability,
signaled by $\mathop{\rm Im} v\ne 0. $  Higher order corrections
help to cure the singularity at $gA_0=\sqrt{gH}, $ which is due to zero
modes in the energy spectrum. For $T<T_{cr}$ there arise a set of regions of
confinement and deconfinement. This is explained by the oscillating
contribution of tachyonic modes to the effective potential.

\section{ Photon PO in U(1)$\times$SU(2) model with
non-abelian vacuum condensate }

Spinor electrodynamics of ``quarks'' with charge $e$ and mass $m$
interacting with SU(2) gauge field is described by the Lagrangian
\be
  {\cal L}=-\frac{1}{4}F_{\mu \nu}F^{\mu \nu}
  -\frac{1}{4g^2}G^a_{\mu\nu}G_a^{\mu\nu}
  +\overline{\psi}(i\gamma^\mu D_\mu-m)\psi
  + e\overline{\psi}\gamma_\mu A^\mu\psi.
\ee
Here $G^a_{\mu\nu}=\partial_\mu G^a_\nu- \partial_\nu G^a_\mu
+f^{abc}G^b_\mu G^c_\nu,\;$ and $D_\mu=\partial_\mu- iG_\mu.$
Photon PO in the momentum representation can be written in the one-loop
approximation as
\be
   \Pi_{\mu\nu}(q,q')=ie^2\delta^4(q+q') \int \! d^4p \:
   \mathop{\rm tr} \left[\gamma_\mu S(p+\frac{q}{2})\gamma_\nu S(p-\frac{q}{2}) \right].
\label{e41}\ee
Here $S(P)=1/(\gamma P-m)$ is the quark Green's function, $P_\mu=p_\mu+G_\mu$.
Performing UV finite integration in (\ref{e41}) the explicit expression for
PO has been obtained. For $q\to 0,$ in the lowest order in $G_\mu$
the antisymmetric part of PO takes the form
$ \displaystyle
  \Pi^A_{\mu\nu}=\frac{5}{6\pi^2} \frac{e^2}{m^2} \mathop{\rm tr}\left[
  G_\mu G_\nu (q_\al G^\al) \right].
$
In the case of the non-abelian spherically symmetric condensate
$G^1_1=G^2_2=G^3_3=\sqrt{\lambda},\;$ $G^a_0=0,\;$
$H^a_i=\delta^a_i \lambda \quad (i=1,2,3)$
the photon topological mass $\Theta_{ind}=\Pi^A(0)\;$
(with $\Pi^A_{\mu\nu}(q)=i\vep_{\mu\nu\al}q^\al \Pi^A(q^2)$)
takes the value~\cite{E_Zh}
$  \displaystyle
  \Theta_{ind}(\lambda)=\frac{5}{24\pi^2}\frac{e^2}{m^2} \lambda^{3/2}.
$
The nonzero antisymmetric part of PO leads to the effect of rotation of
the photon polarization plane.

For large Euclidean momentum $Q^2=-q^2 $ one can improve the well known
perturbative expression for symmetric part of PO by accounting for its
nonanalytical dependence on the background field. To this end the effective
quark mass is introduced: $m_*^2=m^2+3\lambda/4. $
Upon subtracting vacuum contribution one obtains a pure background field
contribution
$  \Pi^S_{\mu\nu}(q) = e^2/(8\pi^2)(Q^2g_{\mu\nu}-Q_\mu Q_\nu)\Pi(Q), $
where
\be
  \Pi(Q)= -4\frac{m^4}{Q^4}  \left(\ln{\left(\frac{m^2}{m_*^2}\right)}+
  \frac{3\lambda}{4m^2}-\frac{\lambda^2}{32m^4} \right).
\ee
This result can be applied to describe radiative corrections to deep
inelastic lepton-hadron scattering $ l(p)+h(P)\to l(p')+X(P') $
with consideration for the model uniform chromomagnetic vacuum field.
Let $q=p-p'$ stand for the momentum transferred. Then the amplitude of this
process is calculated to be $T=T_0 [1-\alpha(Q)\Pi(Q^2)/(2\pi)], $ where
$ \alpha(Q)^{-1}=\alpha^{-1}-1/(3\pi)\ln{(Q^2/m^2)} $ is the fine-structure
constant improved by account for the renormalization group, $T_0$ is the tree
level amplitude. Corresponding correction to the cross-section is
included in the expression
$ d\sigma=d\sigma_0(1-\alpha(Q) \Pi(Q^2) /\pi).$

\end{document}